# JGR Space Physics

## Spatial Variation in the Responses of the Surface External and Induced Magnetic Field to the Solar Wind

**R. M. Shore[1]** [ID]**, M. P. Freeman[1]** [ID]**, J. C. Coxon[2]** [ID]**, E. G. Thomas[3]** [ID]**, J. W. Gjerloev[4,5]** [ID]**, and N. Olsen[6]** [ID]

[1]British Antarctic Survey, Cambridge, UK, [2]Space Environment Physics (SEP) Group, University of Southampton, Southampton, UK, [3]Thayer School of Engineering, Dartmouth College, Hanover, NH, USA, [4]Johns Hopkins University Applied Physics Laboratory, Laurel, MD, USA, [5]Birkeland Center of Excellence, Department of Physics and Technology, University of Bergen, Bergen, Norway, [6]DTU Space, Technical University of Denmark, Kongens Lyngby, Denmark

**Abstract** We analyze the spatial variation in the response of the surface geomagnetic field (or the equivalent ionospheric current) to variations in the solar wind. Specifically, we regress a reanalysis of surface external and induced magnetic field (SEIMF) variations onto measurements of the solar wind. The regression is performed in monthly sets, independently for 559 regularly spaced locations covering the entire northern polar region above 50° magnetic latitude. At each location, we find the lag applied to the solar wind data that maximizes the correlation with the SEIMF. The resulting spatial maps of these independent lags and regression coefficients provide a model of the localized SEIMF response to variations in the solar wind, which we call "Spatial Information from Distributed Exogenous Regression." We find that the lag and regression coefficients vary systematically with ionospheric region, season, and solar wind driver. In the polar cap region the SEIMF is best described by the $B_y$ component of the interplanetary magnetic field (50–75% of total variance explained) at a lag ~20–25 min. Conversely, in the auroral zone the SEIMF is best described by the solar wind $\epsilon$ function (60–80% of total variance explained), with a lag that varies with season and magnetic local time (MLT), from ~15–20 min for dayside and afternoon MLT (except in Oct–Dec) to typically 30–40 min for nightside and morning MLT and even longer (60–65 min) around midnight MLT.

## 1. Introduction

Interaction between the solar wind and the Earth's magnetic field drives a system of electrical currents which couple the magnetosphere and ionosphere in near-Earth space, caused ultimately by disturbances occurring on the Sun (Milan et al., 2017). At ground level, the geomagnetic variations which result from the ionospheric equivalent currents have a well-understood climatology, through the mechanism of the Dungey cycle (Dungey, 1961) and the Expanding-Contracting Polar Cap paradigm (Cowley & Lockwood, 1992; Lockwood & Cowley, 1992; Siscoe & Huang, 1985). Yet, the short-term variability—space weather—of the equivalent currents remains a challenge to predict with better than low accuracy on scales of minutes to hours (Pulkkinen et al., 2013). The terrestrial impacts of unpredictable space weather hazards are broad ranging. Geomagnetically induced currents disrupt and damage large-scale conducting infrastructure, like electricity grids and oil pipelines (Beggan et al., 2013; Knipp, 2015; Pulkkinen et al., 2017). For models of the internal component of the Earth's magnetic field, the external magnetic field "noise" is the single largest source of error (Finlay, Lesur, et al., 2016; Thébault et al., 2017). Improving the predictability of the external geomagnetic variation—to mitigate the space weather hazard—requires both a qualitative and quantitative understanding of how the equivalent currents behave on timescales from minutes to days. A wide selection of modeling methods have been employed to describe solar-terrestrial coupling. Physics-based models (e.g., Vlasiator; von Alfthan et al., 2014) offer the best resolution of theoretical magnetospheric processes, and data assimilation (e.g., Merkin et al., 2016) provides a practical method of linking theory to observed dynamics. However, relating these variations to the equivalent currents observed at ground requires knowledge of the ionospheric conductivity tensor (Forbes, 1981), which is not presently globally measurable and which is known (Laundal et al., 2015; Laundal, Gjerloev, et al., 2016; Laundal, Finlay, et al., 2016; Laundal et al., 2018) to cause strong morphological differences between the true current distribution and the equivalent currents. When indexing geomagnetic activity for the purposes of removing (or reducing) ionospheric and magnetospheric contributions to magnetic field models, it has been common to ignore the morphological differences between ground and space magnetic field measurements. For a recent example, the CHAOS-6 (Finlay, Olsen, et al., 2016) model uses the (ground-based) indices Kp and Dst to select satellite data, and the CM5 (Sabaka et al., 2015) model uses the (space-measured) merging electric field to select ground observatory data. These selections will be more effective in some regions than others, and the lack of accounting for this regional efficacy is a primary cause of the undesired external field signal seen to "leak" into internal field models (e.g., Finlay, Lesur, et al., 2016). We highlight two models which have bridged the description of the magnetic variations in space with those at ground—the model of Weimer (2013) (hereafter W13) and the AMPS model (Laundal, Finlay, et al., 2016; Laundal et al., 2018). In each model the spherical harmonic coefficient amplitudes are given by a regression onto solar wind parameters, thereby resolving the localized effect of the total solar wind direct driving on the ionosphere. In W13 this technique is applied to ground-measured equivalent currents. The AMPS model uses a sophisticated spherical harmonic decomposition of space-measured magnetic variations, to represent the field-aligned currents (FACs) and the horizontal ionospheric current. The divergence-free parts of the latter currents are related to the equivalent current measured at ground (with an assumed ionospheric sheet current altitude). Both AMPS and W13 represent external field variations well on scales of hours and longer, but shorter-period variability is less well described. This is due in part to the temporal averaging applied to the solar wind data on which the models are based—25 min in W13 (Weimer et al., 2010; Weimer, 2013) and 20 min in AMPS. W13 applies an additional time delay of 35 min to the solar wind data, to maximize the correlation between the solar wind and the ionosphere. Since both W13 and AMPS are based on spherical harmonics, these applied latencies are global, with no regional variations. In this study, we demonstrate a technique of regressing equivalent current variations onto measurements of specific solar wind parameters, which is performed independently for each locality. This allows us to explore the localization of two key controlling factors of the solar-terrestrial coupling, which have until now been investigated only in spatially and temporally smoothed terms. First, we assess the geoeffectiveness of specific solar wind drivers, such that their efficacy as indexing parameters for geomagnetic activity is independently quantified for each northern polar location. And second, we investigate the regional distribution of the ionospheric reconfiguration timescales to direct and indirect driving, from those specific solar wind parameters. These timescales will be represented via the most appropriate time delay (lag) between a solar wind perturbation and the peak in the localized ionospheric reconfiguration response. We expect that the terrestrial response for each of these two controlling factors will be modulated by ionospheric conductivity. Hence, we also quantify the variation in these responses as a function of season and solar cycle phase (parameterized by $F10.7$). Where possible, we conduct our investigation by fixing two of these three factors (driver, lag, and conductivity) and varying the third then assessing the impact on the solar wind direct driving geoeffectiveness. To minimize the (smoothing) assumptions, we make about the equivalent current behavior when describing its relationship to the solar wind; we base our investigation on the data set developed by Shore et al. (2018). This is a reanalysis of the Earth's surface external and induced magnetic field (SEIMF) variations, made from the SuperMAG archive (Gjerloev, 2009, 2012). In the next section, we briefly outline how Shore et al. (2018) characterized the component patterns of their reanalysis model (hereafter termed the SEIMF model) and how we will improve upon certain aspects of their techniques. An outcome of our investigation will be a deterministic forecast model for the directly driven parts of the polar ionospheric equivalent current response to the solar wind. We hope this will improve the description of specifically short-timescale ionospheric perturbations. In section 2 we summarize the data from which the model coefficients are calculated. In section 3 we investigate the ionospheric dependence on different solar wind drivers and at different timescales of latency. In section 4 we use the most geoeffective solar wind parameters (determined in section 3) to both investigate the dependence of solar-terrestrial coupling on ionospheric conductivity and to develop a general predictive model for the ionospheric equivalent currents based on the direct solar wind driving. We interpret our findings in a wider context in section 5, and we conclude in section 6.

## 2. Data Description

### 2.1. Reanalysis Data Set of Ionospheric Equivalent Currents

In the Shore et al. (2018) SEIMF reanalysis, the method of data-interpolating empirical orthogonal functions (EOF; modified from Beckers & Rixen, 2003) was used to decompose a given month of magnetic field measurements at 5-min cadence into spatial and temporal patterns of variance. These patterns allowed the SuperMAG data coverage to be "infilled" in space and time, based solely on the spatiotemporal patterns that contributed dominantly to the variability of that month of data. The full SEIMF reanalysis model comprises 144 contiguous monthly EOF analyses, each of which is composed of 10 spatiotemporal variance patterns spanning approximately one calendar month. In order to gain an overview of these 1,440 patterns, Shore et al. (2018) used network analysis (e.g., Caldarelli, 2007) to form "groups" of the variance patterns with shared spatial characteristics. This allowed Shore et al. (2018) to demonstrate that the majority of the variability of the ground-measured magnetic field perturbation can be described in just four dynamical patterns. Shore et al. (2018) sought to contextualize the discovered network analysis groups in terms of direct and indirect solar wind driving. Taking the directly driven response as an example, one must first (subjectively) interpret the grouped EOF patterns then show that these individually conform (largely) to forcing by a given solar wind parameter. In this regard, a particularly useful aspect of the EOF variance decomposition—demonstrated by Shore et al. (2017, 2018)—is that it separates the bulk data into patterns which either correlate more strongly or more weakly with the solar wind driving, when compared to a time series of the original measurements. Shore et al. (2018) compared the four dominant variance patterns to solar wind measurements and (subjectively) demonstrated that three of them represent the well-known disturbance polar ionospheric equivalent current systems DP2, DPY, and DP1. The fourth pattern was interpreted as the variation of the DP2 system according to expansions and contractions of the polar cap and was termed DP2EC. Two of these four major spatiotemporal patterns of variability in the SEIMF (viz., DP2 and DPY) are predictable from the solar wind perturbations. This subjective interpretation was justified further by Shore et al. (2019).

Shore et al. (2018) confirmed the combination of EOF and network analysis as a powerful method in efforts to qualitatively understand the underlying physics of terrestrial magnetic variability and its ionospheric source regions. Yet, there are two main issues with adopting this combined technique for general, quantitative predictions of solar-terrestrial interaction. First, a given solar wind driver may contribute strongly to multiple EOF patterns of variance—no aspect of the EOF decomposition selects either for or against this possibility. Second, there exist intermonthly differences in the variance decomposition of (presumed) directly driven patterns (i.e., DP2 and DPY), which are not clearly related to differences in the direct solar wind driving. The grouping of the EOF patterns via network analysis is insensitive to these two issues. In this way, an unknown quantity of the solar wind driving signal in the reanalysis data remains unexplained by the network analysis grouping, and this amount of lost signal varies between successive months. We aim to address this shortcoming here. The Shore et al. (2018) SEIMF model is presented in the form of spatial and temporal covariance eigenvector pairs (called modes). We take the product of the spatial and temporal parts of each mode to give a reconstruction (in space and time) of part of the variability of the surface-measured magnetic field. We then take the sum of these individual reconstructions for all available modes within each month, plus the monthly mean that was removed from each bin prior to the EOF analysis. This results in the data used in this study, which is a reanalysis data set at 5-min temporal resolution without gaps for 12 continuous years (1997.0–2009.0), defined on a northern polar grid of 559 equal area bins in quasi-dipole (QD) latitude and QD magnetic local time (MLT; Laundal & Gjerloev, 2014; Richmond, 1995). These data are given in the QD component directions $K_{QD}$, vertical; $S_{QD}$, approximately south; $E_{QD}$, approximately east. The SEIMF is often expressed in terms of equivalent currents. In their simplest form, these are 90° rotations of the SEIMF and everywhere proportional. The proportionality could alternatively be chosen to be the infinite plane current which would give that geomagnetic response at ground or a full three-dimensional mathematical derivation of the current system (e.g., Laundal et al., 2018). Here, we adopt the simplest approach which allows a visual interpretation of the external field topology. We accept that this may not apply equally well to the true current distribution in all seasons and locations. Hence, in the supporting information we supply the coefficients of our model in terms of the SEIMF (rather than equivalent current) response to specific solar wind inputs.

### 2.2. Solar Wind Measurements

To represent the solar wind variations, we use OMNI 1-min data (described at http://omniweb.gsfc.nasa.gov/html/HROdocum.html). Specifically, we use the solar wind velocity along the Earth-Sun line ($v_x$), the solar wind proton density ($n_p$), and the north and dawn-dusk components of the interplanetary magnetic field (IMF), respectively IMF $B_z$ and IMF $B_y$ (in the GSM frame; Hapgood, 1992). Each of these parameters is based on measurements already lagged from near the L1 Lagrangian point to their arrival time at the bow shock nose. We apply additional lags to the solar wind data ranging from −10 to +500 min (in 5-min steps up to 150 min then 10-min steps thereafter, totalling 68 separate lags), where a lag of $t - \tau$ for positive $\tau$ indicates solar wind variations at the bow shock preceding the terrestrial response (at a given 5-min mean time $t$) by $\tau$ min. Gjerloev (2012) and Shore et al. (2018) use centered boxcar temporal means to collect the SuperMAG data into 1- and 5-min means, respectively. Hence, the first SEIMF epoch of a given day has a centroid at 2.0 min, based on data spanning epochs −0.5 to 4.5 min. Conversely, the OMNI 1-min data epochs indicate the start of the data used in the average. We collect the OMNI data in 5-min means, so the first set of the day spans 0.0–5.0 min. We regress the 5-min mean OMNI data onto the SEIMF epochs in order to correct for the 0.5-min offset. The regression technique (full details given in the following section) that we use to relate the solar wind variations to the terrestrial magnetic variations only considers linear interactions. To encapsulate nonlinear solar-terrestrial interactions, we add the established solar wind coupling functions $\epsilon$ (Perreault & Akasofu, 1978) and $\Phi_D$ (Milan et al., 2012) to our list of solar wind parameters. The $\epsilon$ parameter is defined as

$$\epsilon = \frac{4\pi}{\mu_0} v_x B^2 \sin^4\left(\frac{\theta_{\text{clock}}}{2}\right) l_0^2 \tag{1}$$

where $\mu_0$ is the permeability of free space $\left(4\pi \times 10^{-7}\right)$; $v_x$ is the (absolute) solar wind velocity in the GSM $X$ direction (Hapgood, 1992); $B$ is the solar wind magnetic field magnitude; $\theta_{\text{clock}}$ is the angle between a projection of the IMF vector onto the GSM $Y$-$Z$ plane and the GSM $Z$ axis, known as the clock angle, that is, $\theta_{\text{clock}} = \tan^{-1}\left(\frac{B_Y}{B_Z}\right)$ (e.g., Milan et al., 2012); and $l_0$ is a length scale factor intended to represent the cross-sectional area over which dayside reconnection takes place—its value here is 7 $R_E$, with $R_E$ the mean Earth radius of 6,371.2 km. Here, $\epsilon$ is given in SI units (e.g., Koskinen & Tanskanen, 2002) and is defined in Watts. From equation (15) in Milan et al. (2012), we have

$$\Phi_D = L_{\text{eff}}\left(v_x\right) v_x B_{YZ} \sin^{\frac{9}{2}}\left(\frac{\theta_{\text{clock}}}{2}\right) \tag{2}$$

where $B_{YZ}$ is the magnitude of the transverse component of the IMF and $L_{\text{eff}}\left(v_x\right) = 3.8 R_E\left(\frac{v_x}{4\times10^5 \text{m/s}}\right)^{\frac{1}{3}}$. Here, $\Phi_D$ is defined in volts. Our motivation for using those particular solar wind coupling functions is that $\epsilon$ is commonly used, and fitting the $\Phi_D$ function to the time derivative of open flux means that this is one of the only coupling functions to physically be defined in terms of the reconnection rate on the dayside (Milan et al., 2012). A wide range of similar one-dimensional solar wind coupling functions have been assessed by Newell et al. (2007). In practice, none of these coupling functions is optimized to describe the equivalent current distribution, and the differences between the responses described by $\epsilon$ and $\Phi_D$ are minimal, as we will later see.

## 3. Investigation 1: Monthly Regressions at a Range of Lags

In this section, we will assess, for each magnetic latitude and MLT locality, which $\tau$ gives the strongest correspondence between each of the solar wind parameters and the terrestrial magnetic variation. We will also assess which of the solar wind parameters best describes each locality.

### 3.1. Method 1: Univariate Linear Regressions With Variable Lags

To compare the solar wind variations to the terrestrial magnetic perturbation, we use a series of independent linear regressions (e.g., Menke, 2012). If $\mathbf{y}$ is a monthlong time series of 5-min reanalysis data from a given magnetic component (say, $S_{\text{QD}}$) at a given location (i.e., a single reanalysis bin in QD latitude and MLT) and $\mathbf{x}_{t-\tau}$ is a time series of a given solar wind parameter lagged by $\tau$ and subsequently sampled to 5-min resolution, then we seek the regression coefficients $u$ and $v$ such that an estimate

$$\hat{\mathbf{y}} = u + v\mathbf{x}_{t-\tau} \tag{3}$$

will minimize $\mathbf{e}^{\mathbf{T}}\mathbf{e}$ where $\mathbf{e} = \hat{\mathbf{y}} - \mathbf{y}$. If we form the model vector $\mathbf{m} = [u, v]^{\mathbf{T}}$, then we can write equation (3) as $\mathbf{y} = \mathbf{Xm} + \mathbf{e}$ or

$$\begin{bmatrix} y_{t_1} \\ y_{t_2} \\ \vdots \\ y_{t_N} \end{bmatrix} = \begin{bmatrix} 1 & x_{t_1-\tau} \\ 1 & x_{t_2-\tau} \\ \vdots & \vdots \\ 1 & x_{t_N-\tau} \end{bmatrix} \begin{bmatrix} u \\ v \end{bmatrix} + \mathbf{e} \tag{4}$$

When forming $\mathbf{X}$ and $\mathbf{y}$, we omit the 5-min epochs for which there was no OMNI data coverage (after lagging and resampling). Over all months, the mean proportion of data removed in this manner is 4.1%, with the set of monthly missing data percentages having a standard deviation of 2.8%. Using the method of least squares (e.g., Menke, 2012), we now solve the problem given in equation (4) for an estimate of the model vector

$$\hat{\mathbf{m}} = (\mathbf{X}^{\mathbf{T}}\mathbf{X})^{-1}\mathbf{X}^{\mathbf{T}}\mathbf{y} \qquad (5)$$

where we assume that the errors $\mathbf{e}$ have zero mean and are uncorrelated with $\mathbf{X}$ and so can be ignored when estimating $\hat{\mathbf{m}}$. Note the distinction between the estimate of the model vector $\hat{\mathbf{m}}$, which signifies that many possible solutions can minimize the least squares norm, and the estimate of the data vector $\hat{\mathbf{y}}$, which shows that there are variations in the data which are not captured by our linear model. In each month, we solve equation (5) independently for each of the 559 latitude-MLT bins, 3 magnetic QD components, 68 different lags $\tau$, and for 7 solar wind parameters, namely, IMF $B_y$, IMF $B_z$ for $|\theta_{\text{clock}}| < 45°$, IMF $B_z$ for $|\theta_{\text{clock}}| \geq 45°$, $v_x$, $n_p$, $\epsilon$, and $\Phi_D$. (The reasoning behind splitting IMF $B_z$ by clock angle is that we expect a different linear behavior for substantially northward IMF.) This set of calculations is repeated separately for each of the 144 months of SEIMF reanalysis data, totalling over 114 million separate linear regressions. We obtain a synoptic map (in QD coordinates) of regression parameters between each solar wind parameter and the terrestrial magnetic variation, for each lag considered, and for each month of data. The synoptic map of $\hat{\mathbf{m}}$ also gives us a linear forecast model of the ground-measured magnetic variations associated with a particular solar wind parameter. We refer to the technique used to produce these synoptic maps as Spatial Information from Distributed Exogenous Regression (SPIDER). For each of the >114 million regressions, we compute (from the same data) the linear Pearson correlation $r$ (Press, 1992) as a measure of the error in our description of $\mathbf{y}$ for a given set of $u$ and $v$. In section 3.2, we present (for some example months) maps of the coefficients $r$ and $v$ and the lags $\tau$ which give the best correlation at each location. We will use these maps to infer the physical causes behind the observed correspondence between the solar wind and the ionospheric equivalent currents. There are three issues with this interpretation: (1) we must assume the FACs flow radially and that the Hall and Pedersen conductivities are uniform in order for the equivalent currents to be morphologically the same as the true (Hall) current distribution (more detail given in the supporting information), (2) correlation does not imply causation, and (3) the different solar wind parameters are not all uncorrelated. Hence, we take the following precautions: first, we only physically interpret the equivalent currents when conductivity is most uniform (summer solstice). Second, we only consider regions of high correlation coefficients to be worth interpreting. Third, we limit our interpretations to regions where the correlation coefficients between the SEIMF and a given solar wind parameter are dominant with respect to correlations with other unrelated solar wind parameters (e.g., IMF $B_y$ and $B_z$). We also refer to our goal of quantifying the efficacy of each solar wind parameter as an activity index. This fully mitigates the difficulties with direct physical interpretation of the equivalent currents, since the maps we present here are simply the geomagnetic variations that will be indexed by a given solar wind driver.

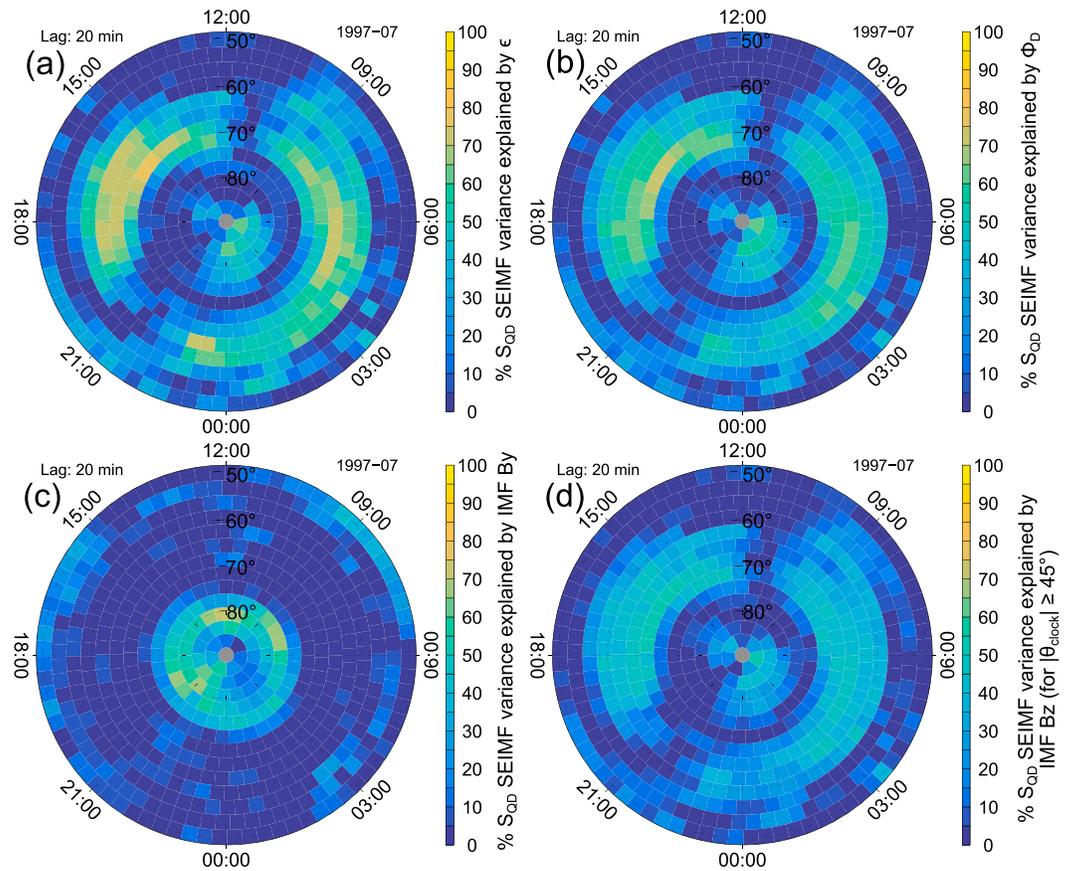

**Figure 1.** For each bin independently, we compute the variance of a linear model prediction, as a percentage of the original data variance, that is, $\left(\frac{\text{var}(\hat{y})}{\text{var}(y)}\right) * 100$, which is $r^2$ expressed as a percentage. Each panel shows the map of the percentage of the $S_{QD}$ component variations explained by linear variations of a different solar wind parameter: (a) $\epsilon$, (b) $\Phi_D$, (c) IMF $B_y$, (d) IMF $B_z$ for $|\theta_{\text{clock}}| \geq 45°$. SEIMF = surface external and induced magnetic field; IMF = interplanetary magnetic field.

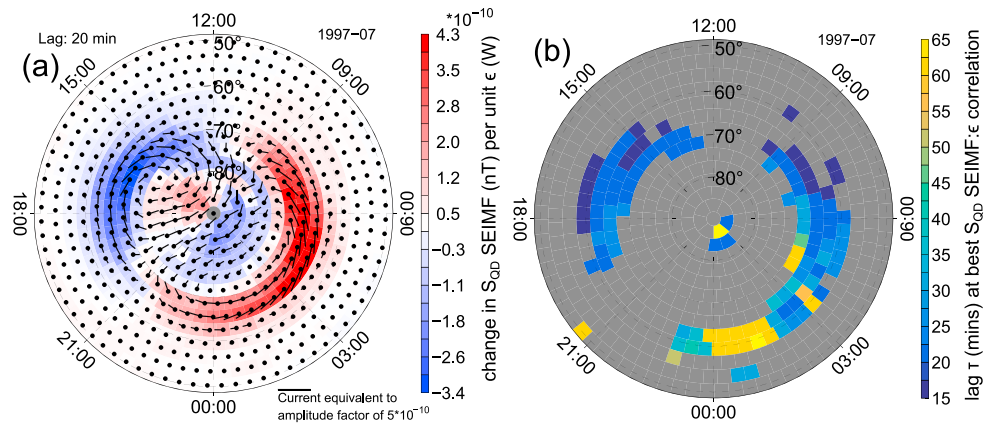

**Figure 2.** (a) Regression coefficients $v$ between the $S_{QD}$ and $E_{QD}$ component quasi-dipole (QD) magnetic variation and the $\epsilon$ parameter lagged by 20 min. The $v$ are shown as vectors rotated by 90° clockwise, to replicate the equivalent current distribution for each unit increase in $\epsilon$. The background colors are for the $S_{QD}$ component (i.e., positive for westward equivalent current). (b) For each spatial bin, the $\tau$ which gave the highest absolute value of $r$ between $S_{QD}$ and $\epsilon$. Bins for which the $r$ value was not in the upper fifth of (considering all lags and all bins) are gray. SEIMF = surface external and induced magnetic field.

### 3.2. Results 1: Ionospheric Equivalent Current Responses to Solar Wind on Varying Timescales

In Figure 1 we show the proportion of $S_{QD}$ component SEIMF variance (from the example month of July 1997) explained in each MLT-latitude bin, by linear predictions from the individual solar wind parameters $\epsilon$, $\Phi_D$, IMF $B_y$, and IMF $B_z$ for $|\theta_{clock}| \geq 45°$. Each linear model in Figure 1 was made using a lag of $\tau = 20$ min, which we will later show is a reasonable estimate for the mean dayside ionospheric reconfiguration timescale due to direct driving.

We highlight that the spatial patterns of explained signal are nearly identical for $\epsilon$, $\Phi_D$, and IMF $B_z$ for $|\theta_{clock}| \geq 45°$, shown, respectively, in Figures 1a, 1b, and 1d. This is expected, since we know from the Dungey cycle and Expanding Contracting Polar Cap (ECPC) paradigms that (southward) IMF $B_z$ is the strongest influence on solar wind-magnetosphere coupling. We also expect the observed behavior that both $\epsilon$ and $\Phi_D$ describe more of the geomagnetic variation than IMF $B_z$, since they incorporate nonlinear solar wind driving. Unexpectedly, we find that $\epsilon$ marginally outperforms $\Phi_D$ in describing the directly driven equivalent current, though this tendency varies for longer $\tau$ reflecting indirect driving (not shown). Taking the proportion of $S_{QD}$ component variance explained by $\epsilon$ (Figure 1a) as an example, we see that these amplitudes peak in the auroral electrojet (AEJ) regions, signifying the expected strong direct driving influence on the electrojets. Conversely, the proportion of variance explained is low in the region of the open-closed field line boundary (OCB) from 70° to 80° latitude and in the convection throats. The throat regions represent meridional plasma flow corresponding to east-west magnetic variations, and indeed, here we observe (not shown) that the $E_{QD}$ component magnetic variations are better described by $\epsilon$. The lower amplitudes of Figure 1a in the OCB region are also expected, since the OCB responds to both direct and indirect driving influences, and our assessment here highlights the direct driving. In comparing Figures 1a and 1c, we find that IMF $B_y$ outperforms $\epsilon$ and $\Phi_D$ in describing the polar cap variability of the $S_{QD}$ component. (The exception to this is in the morning sector—here the $E_{QD}$ component variability (not shown) is stronger and IMF $B_y$ is again its best descriptor). The polar cap is where the azimuthal shear imparted to the open magnetosphere by the solar wind is transferred to the ionosphere, consistent with the models of both Saunders (1989) and Ohtani et al. (1995). The proportions of variance explained by the solar wind parameters $v_x$, $n_p$, and IMF $B_z$ for $|\theta_{clock}| < 45°$ are each weak—we show these in the supporting information. For the rest of this section we focus on the regression model results from the best two descriptors: $\epsilon$, and IMF $B_y$. In Figure 2 we show the outputs from a distributed regression of the reanalysis magnetic field onto the $\epsilon$ parameter, for the same month (July 1997) as in Figure 1. Figure 2a shows the map of regression slope $\nu$ values between $\epsilon$ and the $S_{QD}$ and $E_{QD}$ component QD magnetic variations, expressed as the change in ionospheric equivalent current expected for each unit change in $\epsilon$. This pattern of regression coefficients has the spatial form of the two-cell ionospheric (equivalent) Dungey convection. This confirms our observation from Figure 1 that IMF $B_z$ dominates the ionospheric response to solar wind coupling. Further agreement is offered by the study of Coxon et al. (2019), who have applied the SPIDER technique to data from the Active Magnetosphere and Planetary Electrodynamics Response Experiment. The authors resolved the RR1) and Region 2(R2) FAC distribution in their IMF $B_z$ regression slope maps. Both our findings are consistent with the FACs being driven by azimuthal shears in the same magnetospheric convection flow which underpins the ionospheric equivalent currents (Milan et al., 2017).

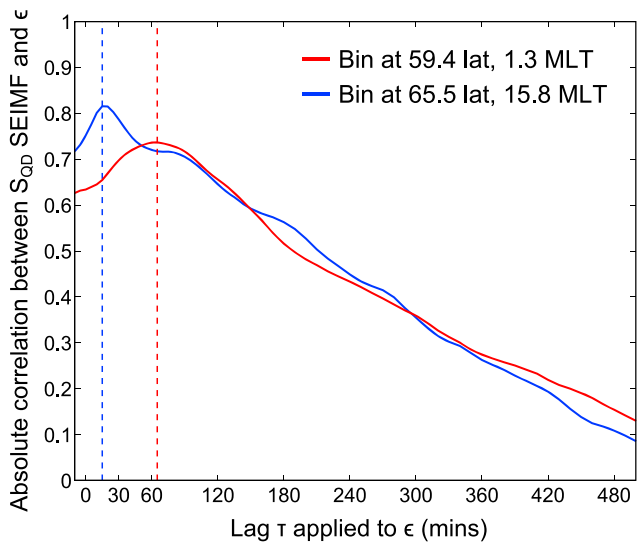

**Figure 3.** Variation in correlation coefficients $r$ between the $S_{QD}$ component magnetic variation and the $\epsilon$ parameter. Shown as a function of lag $\tau$ applied to $\epsilon$, for the two spatial bins in Figure 2b with the fastest (blue) and slowest (red) responses. The vertical dashed lines indicate the lag at which the peak correlation occurred for that location. SEIMF = surface external and induced magnetic field.

In Figure 2b we show the lag $\tau$ in each bin which gives the highest absolute correlation between $\epsilon$ and the $S_{QD}$ component magnetic variations. Values which are not in the upper fifth of the correlations (for all bins and all lags) are shown as gray. We see that the dayside lags in the AEJ have values from 15 to 25 min, with the response time being shorter at lower latitudes and in more sunward sectors. We interpret the poleward increase in the lags to be due to magnetic variations from the OCB—a time-integrated reconnection phenomenon—which sits at the poleward edge of the AEJs (Cowley & Lockwood, 1992; Lockwood & Cowley, 1992; Milan et al., 2017; Siscoe & Huang, 1985). We consider that the decrease in lag toward dayside is because of increased (conjugate) proximity to the dayside magnetopause, where the majority of the interaction described by $\epsilon$ takes place. On the nightside the lags are less coherent, with scattered instances of best correlating lags around 1-hr duration. We interpret these as "overprinting" of the directly driven response by substorms, which will peak in amplitude about 1 hr after a given $\epsilon$ perturbation. These are subjective interpretations. However, it is clear that the spatial distribution of the lags is bimodal, with direct driving limited to lags of 15–25 min and indirect driving having its peak influence about 60-min lag. As summarized by Weimer et al. (2010), a wide range of ionospheric reconfiguration times to solar wind perturbations have been reported in the literature, with peak dayside response timescales ranging from a short as 5 min (Murr & Hughes, 2001) to over 30 min (Blanchard & Baker, 2010). Weimer et al. (2010) themselves used ground magnetometer data split into dayside and nightside sectors and compared these to IMF variations for varying lag. The authors found an ionospheric reconfiguration timescale of 30 min for the dayside and 40–45 min on the nightside. When Weimer et al. (2010) considered all local times at once, they found correlation peaks at lags of 30–35 min. We note that the mean of all the lags we show in Figure 2b is 33 min. The good agreement of this mean value of $\tau$ with Weimer et al.'s (2010) findings leads us to conclude that the Weimer et al. (2010) and Weimer (2013) models are optimized to the mean of a bimodal distribution of $\tau$. We conclude that the ionospheric reconfiguration timescale to solar wind perturbations has a broad range of reported values because there is no single lag which fully describes either the directly or indirectly driven ionospheric equivalent current response. Moreover, given the spatially scattered nature of the $\sim$60-min lags for the sample month shown in Figure 2b, we expect that the best fitting lag at each latitude-MLT location will vary between months and for ionospheric conditions not described by $\epsilon$. In Figure 3 we show the variation in correlation (between the $S_{QD}$ component and $\epsilon$) as a function of $\tau$. These correlation series are shown for the fastest (blue) and slowest (red) responses in Figure 2b. The dashed vertical lines indicate the peak correlation for each series—the entire range of best correlating lags in Figure 2b fits between these two limits of $\tau$. We note that the two series in Figure 3 each has a well-defined peak, implying that variations in the best correlating lag reflect a coherent underlying ionospheric response (this was not clear from Figure 2b alone).

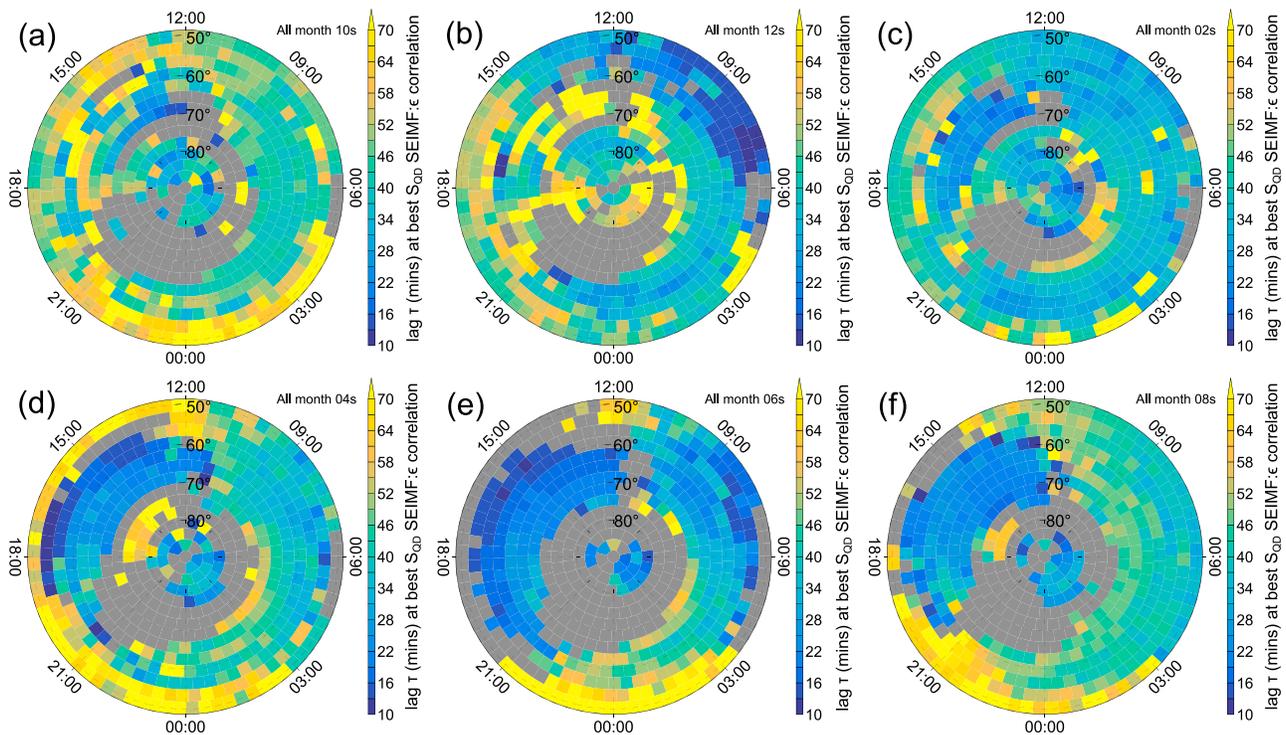

**Figure 4.** The $\tau$ which maximizes correlation between the $S_{QD}$ component variation and $\epsilon$, taken as means which each spans 12 years of the same calendar month. Here we show those means for six sample months: (a) October, (b) December, (c) February, (d) April, (e) June, and (f) August. Only the $\tau$ for the upper fifth of the correlations in each month go into these means; hence, each location's mean is based on a different count of data, and there are some gaps where no high correlations were present. SEIMF = surface external and induced magnetic field.

We now present a systematic overview of the $\tau$ which maximize correlations between $\epsilon$ and the $S_{QD}$ component and how these vary locally with season. We start with the equivalents of the data shown in Figure 2b but for all 144 reanalysis months. These 144 different maps of $\tau$ are split into 12 sets comprising one calendar month each, and the temporal mean is taken for each spatial bin in each set. So we now have one map of best correlating $\tau$ as a mean taken over all 12 Januaries and 11 other months each representing a single calendar month. We show six of these maps in Figure 4, from which there are two main points to make. First, we see that the (duskside) eastward auroral electrojet (EEJ) is characterized by much longer and less spatially coherent lags near winter solstice (Figures 4a–4c) than near summer solstice (Figures 4d–4f). From the monthly mean proportions of $S_{QD}$ variance which is explained by $\epsilon$ (shown in the supporting information Figure S7), we see that the EEJ correlation between $\epsilon$ and the $S_{QD}$ component is likewise lowest at winter solstice. This EEJ region behavior can be understood as follows. The local magnetic field perturbation at ground is given by a spatially integrated combination of the field-aligned, Hall, and Pedersen currents (e.g., Kamide et al., 1981; Laundal et al., 2015; McHenry & Clauer, 1987) plus a noise term from nonionospheric sources. Assuming that the ionospheric electric field is everywhere given by $\epsilon$ (following some lag), then the local contribution from each of these current components is modulated by the ionospheric conductance, itself a combination of insolation, and particle precipitation. In summer, conductance is high and spatially uniform due to the dominance of insolation over precipitation. Then, Fukushima's theorem (Fukushima, 1969; McHenry & Clauer, 1987) applies and the magnetic perturbation is given by the Hall current, with a high signal-to-noise ratio over the nonionospheric sources, and so $S_{QD}$ is closely related to the ionospheric electric field and will correlate well with $\epsilon$. In winter at dusk (i.e., in the EEJ), conductance from both insolation and precipitation is low, so the signal-to-noise ratio is also low, and $S_{QD}$ correlates poorly with $\epsilon$—this explains the lack of coherence in the winter EEJ best correlating $\tau$ between adjacent bins. These case studies are consistent with the findings of Laundal et al. (2015), Laundal, Gjerloev, et al. (2016).

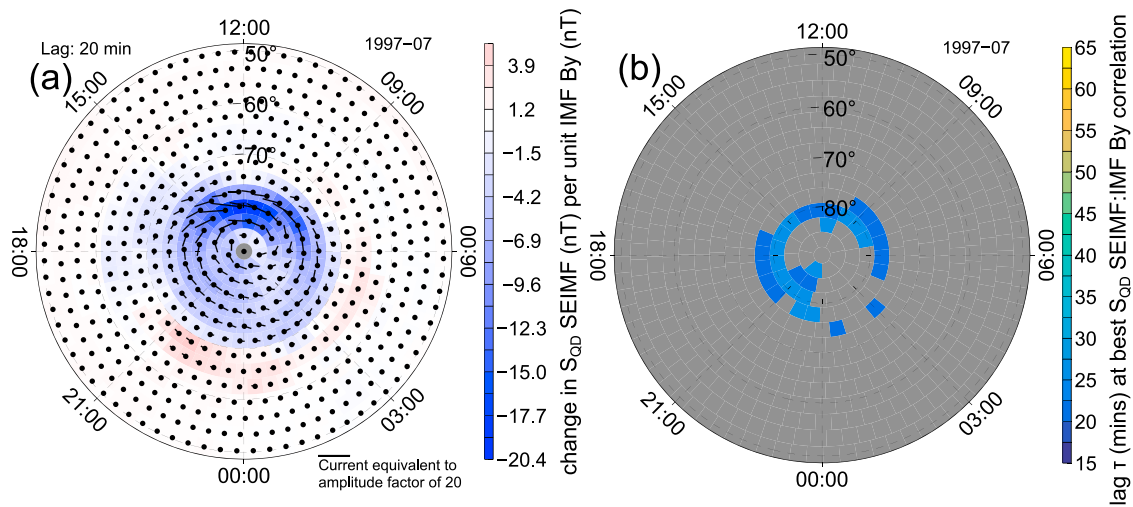

**Figure 5.** As in Figure 2 but for a regression of the geomagnetic field reanalysis onto IMF $B_y$. Note that panel (b) is saturated to the range of values in Figure 2b.

The second main point to discuss from Figure 4 is that the entire dawnside westward auroral electrojet (WEJ) is characterized by longer best correlating $\tau$ than the EEJ. This is consistent for all seasons but is clearest in summer (Figures 4d–4f) due to the more directly driven EEJ then. The WEJ lags are longest in the substorm sector but are longer than the EEJ $\tau$ in all WEJ sectors, persisting even to the noon convection throat (clearest in Figures 4d and 4f). We interpret this to be caused by nightside electron precipitation (possibly from substorms) increasing ionospheric conductivity in patches which are then convected sunward (in the WEJ), until local noon where the plasma will turn antisunward and enter the polar cap (Berkey et al., 1974). Substorms are a time-integrated phenomenon which operate on longer timescales than the direct driving, hence the increased WEJ best correlating $\tau$. The EEJ does not show a similar trend, despite our expectation for higher-energy ions to drift over this sector (Kavanagh et al., 1896), presumably since ion precipitation is less common (Hardy et al., 1989). These SEIMF signals are related to the horizontal currents connecting the R1 and R2 FACs, so we can compare these findings with the results of Coxon et al. (2019). Coxon et al. (2019) find a dawn-dusk asymmetry in the best correlating lags for only the R2 FACs, not the R1 FACs. The authors concluded that the asymmetry was caused by rotation of the currents due to IMF $B_y$ asymmetry during the month studied (March 2010). We cannot yet say whether magnetospheric or ionospheric processes are the dominant controlling factor for the observed dawn-dusk asymmetries in the reconfiguration timescale. Overall, from Figure 4 we show that the EEJ is more directly driven at local summer solstice than at winter solstice and that the WEJ is consistently less directly driven than the summer EEJ. While there is substantial variability in the low-latitude best correlating $\tau$ in Figure 4, this is associated with very low regression coefficient values and may not be physically realistic. After $\epsilon$, the second most important parameter to describe the ionospheric response to the solar wind is IMF $B_y$—we show the results of a regression onto this parameter in Figure 5. Having seen the localization of IMF $B_y$-induced ionospheric shear in Figure 1c, in Figure 5a we can see the sense of this shear, which increases the (QD) eastward equivalent current (i.e., the negative sense of the QD $S_{QD}$ component) by up to 20 nT, per nanotesla increase in (GSM) eastward IMF $B_y$. The magnitude of the IMF $B_y$ effect on the ionosphere peaks in summer solstice—this was also observed by Shore et al. (2018), who did not resolve a DPY pattern in polar winter, implying that IMF $B_y$ forcing in winter has low variability (we confirm this in the supporting information Figure S6). We find a good agreement between the spatial pattern of the SEIMF response to IMF $B_y$ shown in Figure 5a and the FAC response to IMF $B_y$ reported by Coxon et al. (2019). Both results are consistent with FACs bounding a region of azimuthal plasma flow shear on the dayside operating at higher latitudes than the R1 system and an opposing-sense shear region on the nightside according to magnetotail reconnection of skewed field lines.

The set of best correlating $\tau$ for the regression onto IMF $B_y$ is shown in Figure 5b. The range of colors is scaled to that of Figure 2b for easier comparison. This range of $\tau$—within which IMF $B_y$ most directly impacts the ionosphere—fits entirely within the range 20–25 min (following a perturbation at the bow shock nose). These timescales are similar to the range of directly driven timescales reported by Coxon et al. (2019) for the FAC response to IMF $B_y$ driving. For completeness, we show the full range of $\tau$ (i.e., for all correlation values) for both $\epsilon$ and IMF $B_y$ from July 2001 in the supporting information Figure S8. Since the cells in Figure 5b are restricted to the upper fifth of correlations, we can now see the distribution of the best correlating localities more easily than in Figure 1c. The best correlating IMF $B_y$-ionosphere variations are located at the equatorward edge of the polar cap, with two exceptions. First, the correlations at the polar cap edge from 00:00 to 04:00 MLT are lower than elsewhere, possibly reflecting the indirectly driven nature of the WEJ sector. Second, we observe two "spurs" of increased correlation extending poleward at 12:00 and 22:00 MLT. We consider these to represent the pathway taken by newly opened field lines. As discussed by Coxon et al. (2019), this is consistent with the behavior of field lines suggested by both Saunders (1989) and Ohtani et al. (1995). It is remarkable that regression onto IMF $B_y$ (shown in Figure 5) produces an entirely different synoptic map to the regression onto $\epsilon$ for the same month of data (shown in Figure 2). The regression-distribution technique is adept at suppressing irrelevant information, which makes it a powerful tool for physical interpretation. In Figure 1, we showed the importance of selecting the appropriate driver for each location, while in Figures 2–5 we have demonstrated the importance of selecting the most appropriate lag. We now contrast the impact of these two choices on the description of the ionospheric response to solar wind driving. Figure 6a illustrates the difference between the maximum proportion (at each location) of $S_{QD}$ which is explained by $\epsilon$ at any lag and the proportion of $S_{QD}$ which is explained by $\epsilon$ at 20-min lag. In Figure 6b, we likewise show the difference between the proportion of $S_{QD}$ explained by IMF $B_y$ for any lag and for 20-min lag. Both Figures 6a and 6b show that the effect of choosing a location-specific lag is everywhere an improvement (or neutral) in terms of the proportion of geomagnetic variability described by the solar wind function. For $\epsilon$ the improvement is up to 17%, and for IMF $B_y$ it is up to 11% (colocated with regions of best correlating $\tau > 20$ min, shown in Figure S8). In Figure 6c we show the maximum proportion of $S_{QD}$ variability which can be described by selecting either $\epsilon$ or IMF $B_y$, at any lag, for each location. Figure 6d shows the difference between Figures 6c and 1a, demonstrating that the selection of solar wind driver can increase the proportion of $S_{QD}$ variance described by the model by up to 66%. Overall, from Figure 6 we show that selecting the most appropriate solar wind driver (either $\epsilon$ or IMF $B_y$) for each location is the single most important factor in predicting the geomagnetic perturbation. A synthesis of best fitting driver and lag for each location will give the best prediction.

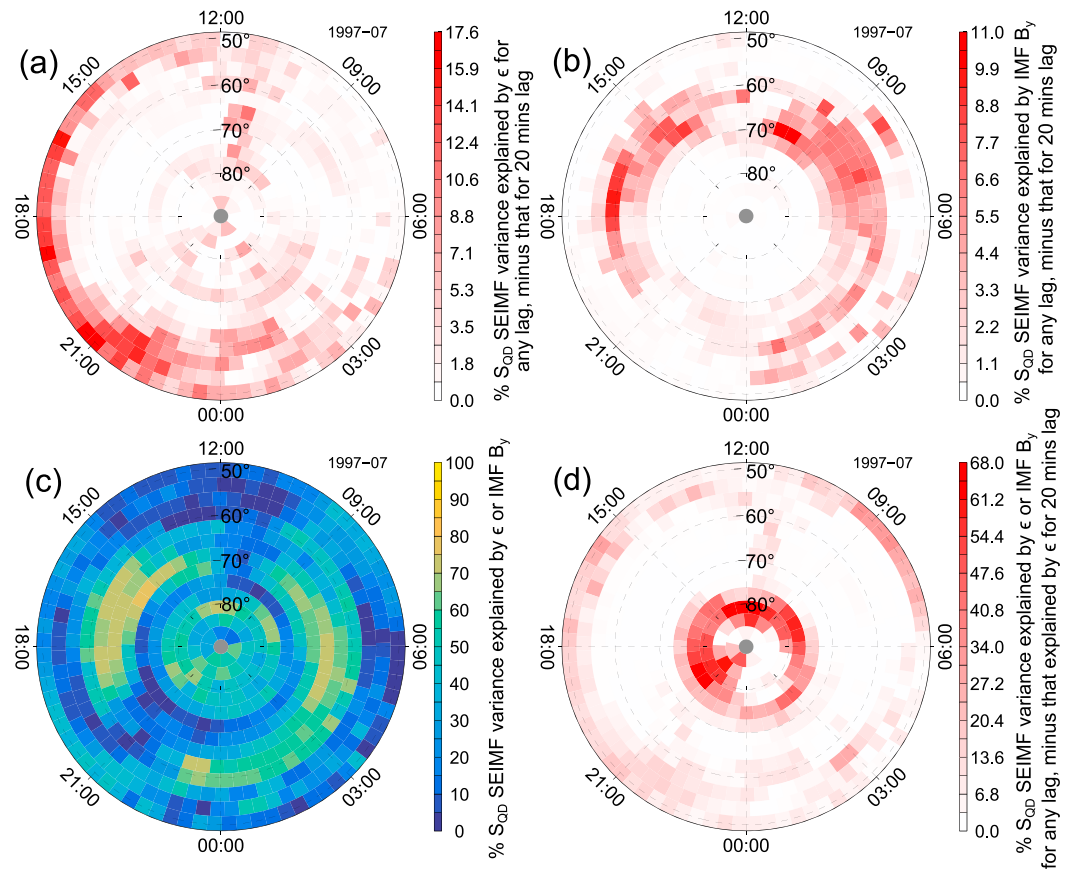

**Figure 6.** (a) Difference map: the maximum proportion of $S_{QD}$ which is explained by $\epsilon$ for any lag, minus the proportion of $S_{QD}$ which is explained by $\epsilon$ for 20-min lag (the latter already shown in Figure 1a). (b) Same as (a) but for IMF $B_y$ instead of $\epsilon$. (c) Proportion of variance explained by choosing the best correlating solar wind parameter (either $\epsilon$ or IMF $B_y$) and the best correlating lag, for each location. (d) Figure 1a minus panel (c). SEIMF = surface external and induced magnetic field. IMF = interplanetary magnetic field.

# 4. Investigation 2: Solar Cycle Length Regressions at Fixed Lag

In section 3 we have focused largely on one sample summer month of Shore et al.'s (2018) 12-year reanalysis. Some seasonal variations were reported in the best correlating $\tau$ between the $S_{QD}$ component and $\epsilon$, thought to be due to variations in insolation. Here we focus on the variations in the strength of the ionospheric response to $\epsilon$ driving at fixed $\tau$ (20 min), as a function of season and solar cycle phase, intended to represent ionospheric conductivity variations. This investigation of conductivity effects mostly assesses differences in insolation, not particle precipitation (as discussed in section 1, the latter is difficult to parameterize). We focus on $\epsilon$ since it was the most important solar wind parameter in defining the ionospheric response, as shown in Figure 1 (mainly because it includes several other solar wind parameters in a nonlinear combination). And we use a fixed lag of 20 min because this was observed in section 3.2 to best represent the directly driven ionospheric variations. A general model of the localized response to direct solar wind driving will result.

## 4.1. Method 2: Conductivity Dependence of the Monthly Regressions, for Fixed Lag

The evaluation of equation (5) for a single solar wind parameter ($\epsilon$) and a fixed $\tau$ was performed independently for all 559 spatial bins, 3 QD components, and 144 months, resulting in 241,488 instances of $\hat{\mathbf{m}}$. For a given single spatial bin and a single QD component, we form a set $\mathbf{a}$ from all 144 regression slope coefficients $v$ (which describe the strength of the localized geomagnetic response to $\epsilon$ driving) and a set $\mathbf{b}$ from all 144 intercepts $u$ (which describe the parts of the geomagnetic variation which are linearly invariant with $\epsilon$). For this single spatial bin and QD component, $\mathbf{a}$ and $\mathbf{b}$ are each one-dimensional vectors of length 144, with one value per month. We now seek a solution for the variation in $\mathbf{a}$ and $\mathbf{b}$ as a multivariate linear function of ionospheric conductivity. We parameterize the ionospheric conductivity with monthly mean $F10.7$ and the sine and cosine of the day of year (corrected to place vernal equinox as day zero, similar to Coxon et al., 2016). We use day of year 79 to represent vernal equinox in each of the 12 years used, but the error resulting from this is not much larger than that resulting from assuming a constant year length of 365.25 days. We refer to a given monthly mean of all 5-min epochs $t$ that went into computing a given $\hat{\mathbf{m}}$, as $t_m$. For simplicity, at that mean epoch we refer to the monthly mean $F10.7$ parameter as $F$, and the sine and cosine terms as $S$ and $C$

$$z = \frac{2\pi \left( t_m - 79 \right)}{365.25} \tag{6}$$

$$S = \sin(z) \tag{7}$$

$$C = \cos(z) \tag{8}$$

If we define $\mathbf{F}$ to be the set of $F$ for all 144 $t_m$ (i.e., a one-dimensional vector of length 144 and likewise for $\mathbf{S}$ and $\mathbf{C}$), then we can pose two multivariate relationships to represent the dependence of $\mathbf{a}$ and $\mathbf{b}$ on conductivity

$$\hat{\mathbf{a}} = c + d\mathbf{F} + e\mathbf{S} + f\mathbf{C} \tag{9}$$

$$\hat{\mathbf{b}} = g + h\mathbf{F} + i\mathbf{S} + j\mathbf{C} \tag{10}$$

where $\hat{\mathbf{a}}$ is an estimate for the variation in $\mathbf{a}$ (and hence is a one-dimensional vector of length 144); likewise $\hat{\mathbf{b}}$ is an estimate of $\mathbf{b}$. The coefficients $e$ and $i$ refer to the solstice variation and $f$ and $j$ refer to the equinox variation. These two are treated separately because we expect the solstice variation to dominate, and together they describe the full annual variation in the data. Now, we redefine our set of independent variables (termed the data kernel matrix in Menke, 2012) $\mathbf{X}$ as

$$\mathbf{X} = \begin{bmatrix} \vec{\mathbf{1}}, \mathbf{F}, \mathbf{S}, \mathbf{C} \end{bmatrix} \tag{11}$$

where $\vec{\mathbf{1}}$ is a vector of ones. If we define a model vector of the coefficients in equation (9) as $\mathbf{m}_a = [c, d, e, f]^\mathsf{T}$, we can solve for an estimate of these parameters via

$$\hat{\mathbf{m}}_a = \left(\mathbf{X}^\mathsf{T}\mathbf{X}\right)^{-1}\mathbf{X}^\mathsf{T}\mathbf{a} \tag{12}$$

which minimizes the misfit of $\hat{\mathbf{a}} - \mathbf{a}$. Likewise, we solve for estimates of $g$, $h$, $i$, and $j$ which best fit $\mathbf{b}$. Now, we have the general model to predict, at a given epoch $t$ the geomagnetic perturbation

$$\hat{y_{est}} = \hat{b} + \hat{a}\epsilon_{t-20} \tag{13}$$

where $\hat{a}$ is an scalar estimate of $\mathbf{a}$ at a single epoch $t$ (likewise for $\hat{b}$). We distinguish between $\hat{y_{est}}$ and $\hat{y}$ since we do not explicitly minimize the misfit of $\mathbf{y_{est}^{\hat{}}} - \mathbf{y}$ for all months.

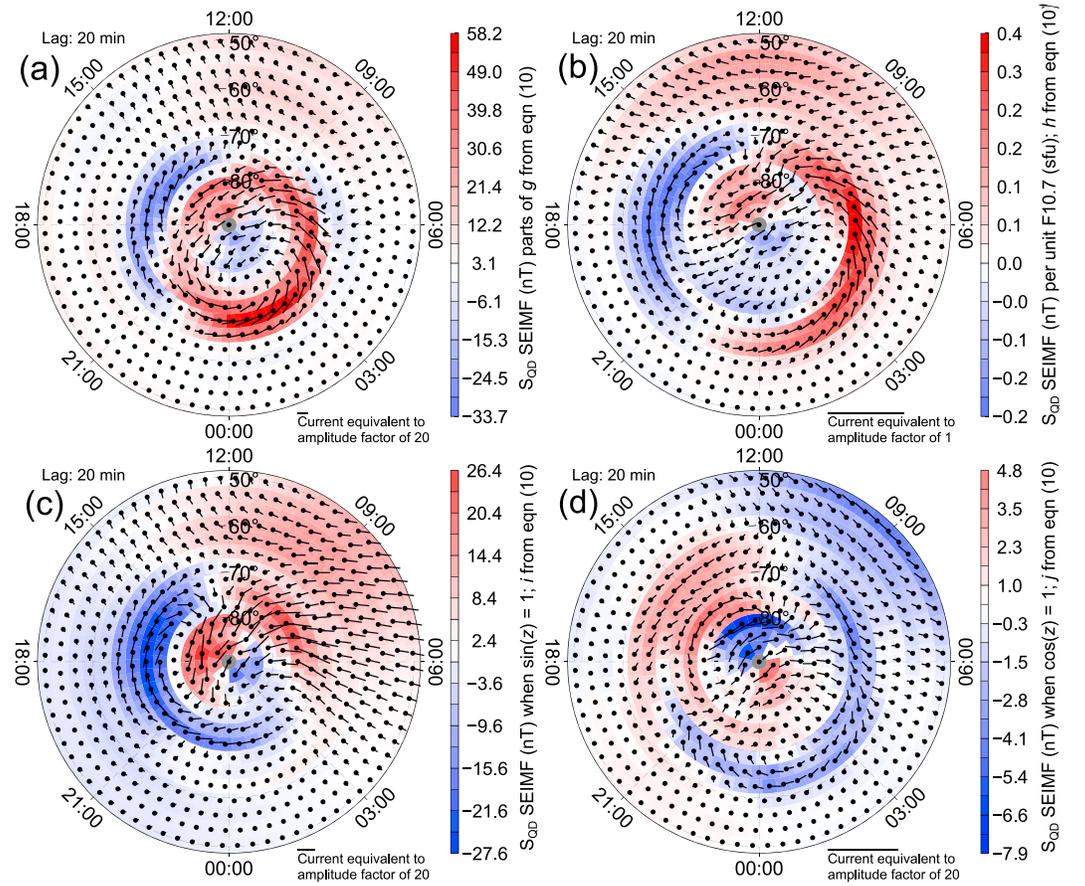

**Figure 7.** Spatial distribution of the regression coefficients from equation (10), shown as equivalent horizontal current vectors (e.g., as in Figure 2a). The panels show maps of the geomagnetic variations which are independent of $\epsilon$, which are divided into the following components: (a) $g$, which is independent of conductivity variations; (b) $h$, which varies linearly with $F10.7$; (c) $i$, which varies according to the sine of day of year from vernal equinox; and (d) $j$, which varies according to the cosine of day of year from vernal equinox. SEIMF = surface external and induced magnetic field.

The multivariate regression described by equation (12) is applied to $\mathbf{a}$ and $\mathbf{b}$ for each of the 559 bins and 3 QD components independently. The result is a forecast model of the equivalent current (or the geomagnetic perturbation) 20 min in the future, given a contemporaneous measurement of $\epsilon$. Given the eight conductivity parameters (including intercepts) at each latitude-MLT bin and for each QD component, our model has 13,146 parameters, which we supply in the supporting information. We assume with equations (9), (10), and (13) that the conductivity at a given location varies linearly with $F$, while $S$ gives the difference between solstices, and $C$ gives the differences (if any) between equinoxes. The unmodeled parts of the geomagnetic variation will be those which vary nonlinearly with $F10.7$ or $\epsilon$ or which vary nonsinusoidally with day of year. In the supporting information (Figure S9), we show the proportion of **b** described by **b̂**, and also the proportion of **a** described by **â**, showing that the majority of the full variability in the 144 month sets of model coefficients is represented in our model. We expect that this proportion would increase with a more complete (i.e., not just $\epsilon$ based) description of the geomagnetic variations in each month.

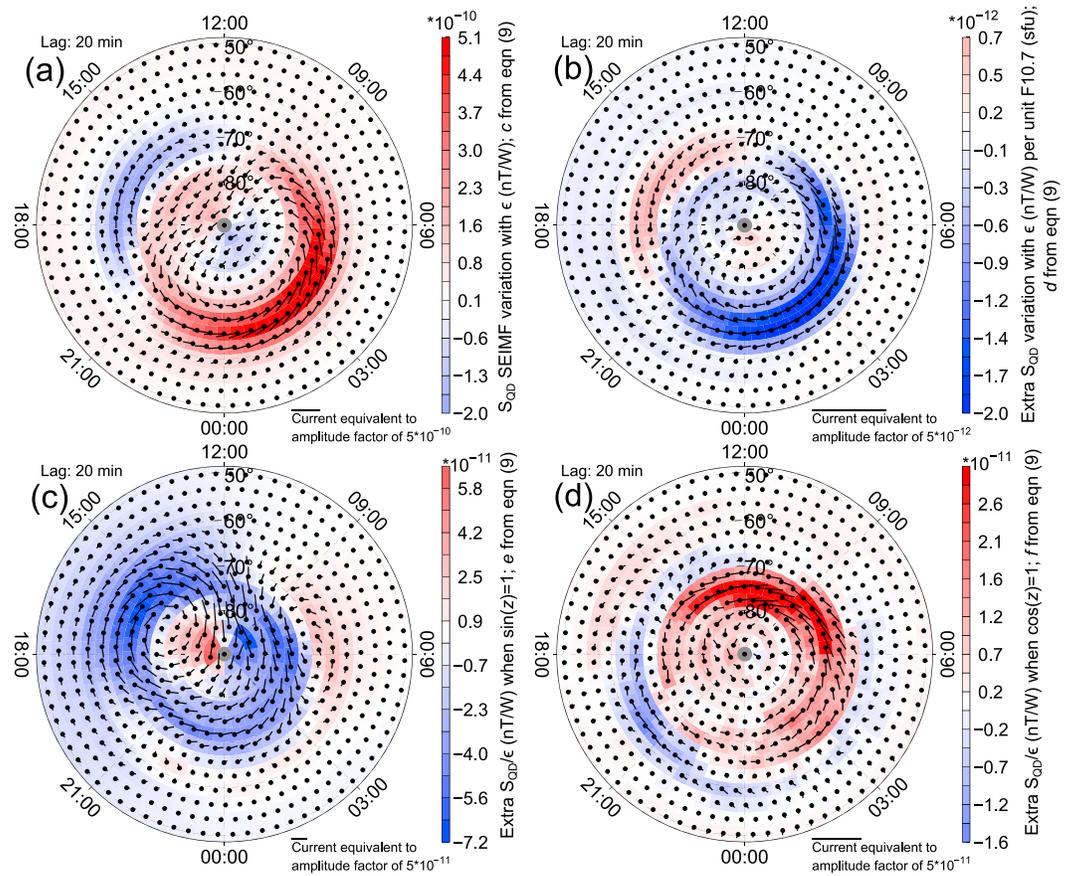

**Figure 8.** Spatial distribution of the regression coefficients from equation (9), shown as equivalent horizontal current vectors. The panels show maps of the dependence of the geomagnetic variation on a unit change in $\epsilon$, divided into the following components: (a) $c$, which is the geomagnetic dependence on $\epsilon$ which operates independently of season or solar cycle phase; (b) $d$, which is the modulation of the geomagnetic dependence on $\epsilon$ for each unit change in $F10.7$; (c) $e$, which is the modulation of the geomagnetic dependence on $\epsilon$ according to the sine of day of year from vernal equinox; and (d) $f$, which is the modulation of the geomagnetic dependence on $\epsilon$ according to the cosine of day of year from vernal equinox. SEIMF = surface external and induced magnetic field.

### 4.2. Results 2: Conductivity Dependence of Solar Wind-Ionosphere Coupling

The set **b** comprises the parts of the geomagnetic variation which operate independently of reconnection rate, since it is a collection of intercept coefficients from independent monthly regressions of the SEIMF onto $\epsilon$. It is thus possible that **b** contains the ionospheric representation of the viscous interaction between the solar wind and the magnetosphere (Axford, 1964). However, **b** will also contain inputs from IMF $B_y$, Sq, positive IMF $B_z$, substorms, and the ground-induced components of all these, none of which is separated out by this regression. In Figure 7, we show the conductivity dependence of this ensemble of reconnection-independent variations, in the form of equivalent horizontal current vector maps of the regression coefficients from equation (10). Figure 7a shows the map of $g$—these are the parts of the external geomagnetic field which vary independently of $\epsilon$ and which have no dependence on $F10.7$ or season.

The signal is strongest near the midnight sector of the WEJ, implying a substorm contribution. The overall two-cell nature of the pattern may indicate the aforementioned viscous interaction.

Figure 7b shows the map of $h$, indicating that an increase in $F10.7$ will increase the (reconnection rate independent) geomagnetic variation, predominantly on the dayside. This takes the form of increased two-cell convection and an increased Sq current, the latter strongest at noon and equatorward of 60° latitude. Figures 7c and 7d show the impact of season on the $\epsilon$-independent magnetic variation. Differences in solstice have the largest impact, acting to increase the duskside convection cell amplitude. Note also the alteration (in both Figures 7c and 7d) of equivalent current at low latitudes equatorward of 75° colatitude in the sector from 06:00 to 12:00 MLT. The variation in this region is strongest according to season; hence, it could be due to neutral wind variation (which is modulated by tilt angle; e.g., Billett et al., 2018) or due to the faster buildup of conductivity in daylight hours during summer. In Figure 8 we show the maps of the regression coefficients from equation (9)—these are similar results to those shown in Figure 7 but for the parts of the geomagnetic variation which are dependent on dayside reconnection rate. Figure 8a shows the map of $c$—this is the part of the geomagnetic field's dependence on $\epsilon$ which operates independent of conductivity variations, which resembles the two-cell convection. It is modulated by the map of $d$ shown in Figure 8b. We see that the effect of increasing $F10.7$ is to uniformly decrease the dependence of the two-cell convection on $\epsilon$. Weimer (2013) noted that greater magnetic perturbations sometimes occur with lower $F10.7$ but did not suggest why. We are also unable to determine the underlying cause of this observation, but we highlight the distinction between Figures 7b and 8b, demonstrating that increasing $F10.7$ acts to increase the $\epsilon$-independent geomagnetic variation, while decreasing its dependence on $\epsilon$. In other words, equivalent values for $\epsilon$ result in greater magnetic perturbations when $F10.7$ is lower.

In Figure 8c we see that solstice is the dominant control on $\epsilon$ dependence of the geomagnetic variation. The effect of going from winter to summer is to append the majority of the duskside equivalent convection cell amplitude. This is in good agreement with the resolution of seasonal effects in the two-cell equivalent convection pattern's variability by Shore et al. (2018) and the finding of Laundal et al. (2018) that the WEJ is the only large-scale current system that exists in all seasons. We also note that the solstice variation in Figure 8c extends to rather low (certainly subauroral) latitudes between MLTs of 13:00 and 22:00 (i.e., the duskside). We consider that this is related to the trends discussed above in conjunction with Figures 4 and S7. Specifically, we suggest that in winter, when the EEJ conductivity is low, the conductivity gradient between auroral latitudes and lower latitudes is also lowered. This could allow the R2 FAC to extend to lower latitudes in winter at dusk, which would increase the dependence of $S_{QD}$ upon $\epsilon$ at subauroral latitudes. (This effect can also be seen by comparing the duskside subauroral regions of Figures S7b and S7e, below 65°, where the proportion of $S_{QD}$ explained by $\epsilon$ is highest in winter). In Figure 8d we show the alteration of $\epsilon$ dependence as we change from autumn to spring equinox. The effect is half the magnitude of the impact from changing solstice and has the same spatial pattern as Figure 5a. This could imply that positive IMF $B_y$ has a stronger effect on the ionosphere in autumn but is more likely to signify a lateral shift of the entire two-cell convection pattern in the dawn-dusk meridian. Note that it is not clear whether our specification of equinox (day of year 79) for $S$ and $C$ in equations (7) and (8) will perfectly separate solsticial and equinoctial variations. For instance, Russell and McPherron (1973) noted an offset between the annual variation in the expected solar wind driving and the dates of seasons defined by daylight. Hence, we cannot rule out the possibility that the variations in Figure 8d are in part, solsticial variations. And Shore et al. (2018) have already noted a solsticial difference in the (purported) convection cell position in the predawn meridian (their Figure 6) which could be explained by the pattern in Figure 8d.

## 5. Discussion

In contrast to Shore et al. (2017, 2018, 2019) who used time series of spatial patterns presumed (without direct evidence) to represent the ionospheric response to solar wind driving, here we have resolved the "true" spatial signal of the linear equivalent current response to specific solar wind drivers. From section 3 we see that our distributed regression "SPIDER" technique is adept at extracting this synoptic information from a mass of overlapping signals. This technique allows us to assess how much of the equivalent current at each MLT-latitude locality is related to a given solar wind driver and hence to quantify that driver's geoeffectiveness at each specific location. This has led to the two main results of this report: First that at different MLT-latitude locations, a different IMF component may be the best predictor of the equivalent currents. Second, different temporal lags of these components will be the best predictor at different locations. As an extension of these findings, we show that all "universal" solar wind coupling functions (e.g., Milan et al., 2012; Newell et al., 2007) have local variations in how effectively they describe the magnetic perturbations driven by solar-terrestrial coupling. In other words, these coupling functions should not be considered universal. While this study focuses on directly driven signals, we have also resolved the statistical signatures of indirectly driven responses at longer lags, such as the equivalent current dependence on the OCB location and the substorm contribution (i.e., the DP2EC and DP1 groups from Shore et al., 2018). However, there is no well-defined nightside equivalent for measurements of the dayside reconnection rate, and hence, we have not been able to isolate the substorm response as an independent phenomenon here. Nightside reconnection depends on time-integrated effects with poorly understood mechanisms, which are not presently predictable nor directly measured. Investigation of these mechanisms is beyond the scope of this particular study. We recommend that the Shore et al. (2018) model be used when the description of specific indirect driving effects is important. The SPIDER technique (as presented here) is best used when forecast or nowcast of specific directly driven responses is desired. The closest comparison we can make is to the W13 empirical forecast model of Weimer (2013), described in section 1. The W13 model encapsulates nonlinear solar wind driving and thus will describe a larger proportion of the full ionospheric variability than the general forecast model we present in section 4. However, since the W13 model is based on quasi-global harmonic functions, it cannot utilize localized $\tau$. We consider that the resulting smoothing between direct and indirect driving processes is one cause for its loss of forecast accuracy on short timescales, reported by Weimer (2013). We also highlight the study of Dods et al. (2017), who have used network analysis of ground-based magnetometer data to resolve the timescales of the ionospheric response to IMF north-south turnings. This study is complementary to our results and technique. Dods et al. (2017) presented a finer temporal resolution (2 min) for the response to IMF perturbations to that shown here. The advantages of our study are first that we can resolve the ionospheric response to any solar wind state, rather than a superimposed set of specific events (e.g., IMF turnings). Second, with our technique we can resolve both the spatiotemporal magnetic perturbation pattern and the timescales it operates over, while the technique of Dods et al. (2017) gives the latter of these two. The forecast model presented in section 4 is a proof of concept, intended to demonstrate the impact of ionospheric conductivity variations on the geoeffectiveness of the $\epsilon$ direct driving. We note several shortcomings of this general forecast model, which will need to be overcome to make its prediction more accurate than the Weimer (2013) model. Since our model is based solely on $\epsilon$, it does not yet account for the azimuthal asymmetries imposed by IMF $B_y$ nor does it represent the driving for northward IMF $B_z$. Despite our own results showing that the use of a single global latency is inadequate, we have based the forecast model on a fixed $\tau$ of 20 min. This is justified, since we aim to represent just the direct driving, and we have observed that a single "best" $\tau$ cannot be picked to represent indirectly driven effects, particularly for nightside localities. Taking account of the separate timescales of the indirect driving, the persistence of the solar wind and the localized memory of the ionospheric response are not in the scope of this investigation. Lastly, our model also does not presently resolve nonlinear driving from the solar wind, with the exception of the already nonlinear formulation of $\epsilon$ and the sine and cosine of day of year. Nonlinear ionospheric responses to the solar wind are likely common, for instance, when the solar wind suddenly changes speed or density, or when an IMF component changes sign, or when a given solar wind parameter has extremely high or low amplitude. All of these shortcomings in our proof-of-concept model can be ameliorated within the SPIDER technique framework.

## 6. Conclusions

We present a new technique for extracting synoptic overviews of solar-terrestrial coupling, from large data sets of terrestrial magnetic field data. The technique involves regressing monthly time series of geomagnetic field variations from individual latitude-MLT locations onto measurements of solar wind parameters. Despite the regressions being independently performed for each locality, we resolve a high degree of inter-location coherence. This allows us to use the distribution of regression parameters as a synoptic map, which defines the spatial distribution of the ionospheric response to a given solar wind parameter. We call this technique SPIDER. Since the regression is performed individually for each locality, we are able to resolve different latencies between each solar wind parameter (as represented at the bow shock nose) and the peak ionospheric equivalent current response (measured on ground). Our description of the ionospheric reconfiguration timescale has the highest spatial resolution to date. We reveal a systematic dawn-dusk difference in the ionospheric reconfiguration timescale for the EEJ and WEJ. We also find that a given solar wind driver will describe different amounts of the equivalent current response in different locations. In other words, different locations are not equally well predicted by the same solar wind state. This predictability is modulated by season and $F10.7$. For instance, we find a seasonal variation in the predictability of the eastward electrojet, and we find that increased $F10.7$ acts to decrease the dependence of the geomagnetic variation on $\epsilon$ driving. We contrast the importance of the solar wind driver, and the lag applied to it, in describing the ionospheric equivalent current response. We find that the specific driver is generally a more important factor in obtaining the best prediction, although lag is important to understand the underlying physical processes. Furthermore, we show that the best prediction of the ionospheric response would be gained from an ensemble of contributions at a range of lags and that the form of the ensemble will vary with location.

**Acknowledgments**
This work was funded by the Natural Environment Research Council under Grant NE/J020796/1 and uses datasets funded by Grant NE/J020796/1. The work was performed using hardware and support at the British Antarctic Survey (BAS). The SuperMAG data were obtained directly from Jesper Gjerloev but can be accessed online (http://supermag.jhuapl.edu/). The ACE data (IMF, solar wind velocity, and proton density) were obtained from ftp://spdf.gsfc.nasa.gov/pub/data/omni/high_res_omni/monthly_1min/ on 3 June 2014. $F10.7$ data were obtained from ftp://ftp.ngdc.noaa.gov/STP/GEOMAGNETIC_DATA/INDICES/KP_AP/ on 19 December 2012. The EOF reanalysis is available in the supporting information of Shore et al. (2018). We would like to thank Gareth Chisham and Mike Lockwood for helpful discussions. R. M. S. and J. C. C. would like to thank the organizers and sponsors of the Russian-British Seminar of Young Scientists on "Dynamical plasma processes in the heliosphere: from the Sun to the Earth," which hosted discussions instrumental to the collaboration on this study. For the ground magnetometer data we gratefully acknowledge INTERMAGNET (we thank the National Institutes that support its contributing magnetic observatories and INTERMAGNET for promoting high standards of magnetic observatory practice [www.intermagnet.org]); USGS, Jeffrey J. Love; CARISMA, PI Ian Mann; CANMOS; The S-RAMP Database, PI K. Yumoto and Dr. K. Shiokawa; The SPIDR database; AARI, PI Oleg Troshichev; The MACCS program, PI M. Engebretson, Geomagnetism Unit of the Geological Survey of Canada; GIMA; MEASURE, UCLA IGPP and Florida Institute of Technology; SAMBA, PI Eftyhia Zesta; 210 Chain, PI K. Yumoto; SAMNET, PI Farideh Honary; The institutes who maintain the IMAGE magnetometer array, PI Eija Tanskanen; PENGUIN; AUTUMN, PI Martin Connors; DTU Space, PI Dr. Anna Naemi Willer; South Pole and McMurdo Magnetometer, PI's Louis J. Lanzarotti and Alan T. Weatherwax; ICESTAR; RAPIDMAG; PENGUIn; British Antarctic Survey; McMac, PI Dr. Peter Chi; BGS, PI Dr. Susan Macmillan; Pushkov Institute of Terrestrial Magnetism, Ionosphere and Radio Wave Propagation (IZMIRAN); GFZ, PI Dr. Juergen Matzka; MFGI, PI B. Heilig; IGFPAS, PI J. Reda; University of L'Aquila, PI M. Vellante; and SuperMAG, PI Jesper W. Gjerloev.